\documentclass[aps,
prd,
reprint,
preprintnumbers,
superscriptaddress,
floatfix]{revtex4-1}

\usepackage{graphicx}
\usepackage{xspace}
\usepackage{units}
\usepackage{hyperref}
\usepackage{lineno}
\usepackage{verbatim}
\usepackage[dvipsnames]{xcolor}

\usepackage{float}

\hypersetup{hidelinks}

\newcommand{\figurescale}{0.99\columnwidth}

\def\nova{NOvA\xspace}

\begin{document}

\preprint{FERMILAB-PUB-26-0496-PPD}

\title{Timing-Based Search for Magnetic Monopoles with the NOvA Detector on the Surface}

\newcommand{\ANL}{Argonne National Laboratory, Argonne, Illinois 60439, 
USA}
\newcommand{\Bandirma}{Bandirma Onyedi Eyl\"ul University, Faculty of 
Engineering and Natural Sciences, Engineering Sciences Department, 
10200, Bandirma, Balıkesir, Turkey}
\newcommand{\ICS}{Institute of Computer Science, The Czech 
Academy of Sciences, 
182 07 Prague, Czech Republic}
\newcommand{\IOP}{Institute of Physics, The Czech 
Academy of Sciences, 
182 21 Prague, Czech Republic}
\newcommand{\Atlantico}{Universidad del Atlantico,
Carrera 30 No.\ 8-49, Puerto Colombia, Atlantico, Colombia}
\newcommand{\BHU}{Department of Physics, Institute of Science, Banaras 
Hindu University, Varanasi, 221 005, India}
\newcommand{\UCLA}{Physics and Astronomy Department, UCLA, Box 951547, Los 
Angeles, California 90095-1547, USA}
\newcommand{\Caltech}{California Institute of 
Technology, Pasadena, California 91125, USA}
\newcommand{\CUSB}{Central University of South Bihar,
NH-120, Gaya Panchanpur Road, Post Fatehpur, 
Gaya, Bihar, 824 236, India}
\newcommand{\Charles}
{Charles University, Faculty of Mathematics and Physics,
 Institute of Particle and Nuclear Physics, Prague, Czech Republic}
\newcommand{\Cincinnati}{Department of Physics, University of Cincinnati, 
Cincinnati, Ohio 45221, USA}
\newcommand{\CSU}{Department of Physics, Colorado 
State University, Fort Collins, CO 80523-1875, USA}
\newcommand{\CTU}{Czech Technical University in Prague,
Brehova 7, 115 19 Prague 1, Czech Republic}
\newcommand{\Dallas}{Physics Department, University of Texas at Dallas,
800 W. Campbell Rd. Richardson, Texas 75083-0688, USA}
\newcommand{\DallasU}{University of Dallas, 1845 E 
Northgate Drive, Irving, Texas 75062 USA}
\newcommand{\Delhi}{Department of Physics and Astrophysics, University of 
Delhi, Delhi 110007, India}
\newcommand{\JINR}{Joint Institute for Nuclear Research,  
Dubna, Moscow region 141980, Russia}
\newcommand{\Erciyes}{
Department of Physics, Erciyes University, Kayseri 38030, Turkey}
\newcommand{\FNAL}{Fermi National Accelerator Laboratory, Batavia, 
Illinois 60510, USA}
\newcommand{\FSU}{Florida State University, Tallahassee, Florida 32306, USA}
\newcommand{\UFG}{Instituto de F\'{i}sica, Universidade Federal de 
Goi\'{a}s, Goi\^{a}nia, Goi\'{a}s, 74690-900, Brazil}
\newcommand{\Guwahati}{Department of Physics, IIT Guwahati, Guwahati, 781 
039, India}
\newcommand{\Harvard}{Department of Physics, Harvard University, 
Cambridge, Massachusetts 02138, USA}
\newcommand{\Homi}{Homi Bhabha National Institute, Training School Complex,
 Anushakti Nagar, Mumbai 400094, India}
\newcommand{\Houston}{Department of Physics, 
University of Houston, Houston, Texas 77204, USA}
\newcommand{\IHyderabad}{Department of Physics, IIT Hyderabad, Hyderabad, 
502 205, India}
\newcommand{\Hyderabad}{School of Physics, University of Hyderabad, 
Hyderabad, 500 046, India}
\newcommand{\IIT}{Illinois Institute of Technology,
Chicago IL 60616, USA}
\newcommand{\Imperial}{Imperial College London, Department of Physics,
 London, United Kingdom}
\newcommand{\Indiana}{Indiana University, Bloomington, Indiana 47405, 
USA}
\newcommand{\INR}{Institute for Nuclear Research of Russia, Academy of 
Sciences 7a, 60th October Anniversary prospect, Moscow 117312, Russia}
\newcommand{\UIowa}{Department of Physics and Astronomy, University of Iowa, 
Iowa City, Iowa 52242, USA}
\newcommand{\ISU}{Department of Physics and Astronomy, Iowa State 
University, Ames, Iowa 50011, USA}
\newcommand{\Irvine}{Department of Physics and Astronomy, 
University of California at Irvine, Irvine, California 92697, USA}
\newcommand{\Jammu}{Department of Physics and Electronics, University of 
Jammu, Jammu Tawi, 180 006, Jammu and Kashmir, India}
\newcommand{\Lebedev}{Nuclear Physics and Astrophysics Division, Lebedev 
Physical 
Institute, Leninsky Prospect 53, 119991 Moscow, Russia}
\newcommand{\Magdalena}{Universidad del Magdalena, Carrera 32 No 22-08 Santa Marta, Colombia}
\newcommand{\MSU}{Department of Physics and Astronomy, Michigan State 
University, East Lansing, Michigan 48824, USA}
\newcommand{\Crookston}{Math, Science and Technology Department, University 
of Minnesota Crookston, Crookston, Minnesota 56716, USA}
\newcommand{\Duluth}{Department of Physics and Astronomy, 
University of Minnesota Duluth, Duluth, Minnesota 55812, USA}
\newcommand{\Minnesota}{School of Physics and Astronomy, University of 
Minnesota Twin Cities, Minneapolis, Minnesota 55455, USA}
\newcommand{\Mississippi}{University of Mississippi, University, Mississippi 38677, USA}
\newcommand{\NISER}{National Institute of Science Education and Research, 
An OCC of Homi Bhabha
National Institute, Bhubaneswar, Odisha, India}
\newcommand{\OSU}{Department of Physics, Ohio State University, Columbus,
Ohio 43210, USA}
\newcommand{\Oxford}{Subdepartment of Particle Physics, 
University of Oxford, Oxford OX1 3RH, United Kingdom}
\newcommand{\Panjab}{Department of Physics, Panjab University, 
Chandigarh, 160 014, India}
\newcommand{\Pitt}{Department of Physics, 
University of Pittsburgh, Pittsburgh, Pennsylvania 15260, USA}
\newcommand{\QMU}{Particle Physics Research Centre, 
Department of Physics and Astronomy,
Queen Mary University of London,
London E1 4NS, United Kingdom}
\newcommand{\RAL}{Rutherford Appleton Laboratory, Science 
and 
Technology Facilities Council, Didcot, OX11 0QX, United Kingdom}
\newcommand{\SAlabama}{Department of Physics, University of 
South Alabama, Mobile, Alabama 36688, USA} 
\newcommand{\Carolina}{Department of Physics and Astronomy, University of 
South Carolina, Columbia, South Carolina 29208, USA}
\newcommand{\SDakota}{South Dakota School of Mines and Technology, Rapid 
City, South Dakota 57701, USA}
\newcommand{\SMU}{Department of Physics, Southern Methodist University, 
Dallas, Texas 75275, USA}
\newcommand{\Stanford}{Department of Physics, Stanford University, 
Stanford, California 94305, USA}
\newcommand{\Sussex}{Department of Physics and Astronomy, University of 
Sussex, Falmer, Brighton BN1 9QH, United Kingdom}
\newcommand{\Syracuse}{Department of Physics, Syracuse University,
Syracuse NY 13210, USA}
\newcommand{\Tennessee}{Department of Physics and Astronomy, 
University of Tennessee, Knoxville, Tennessee 37996, USA}
\newcommand{\Texas}{Department of Physics, University of Texas at Austin, 
Austin, Texas 78712, USA}
\newcommand{\Tufts}{Department of Physics and Astronomy, Tufts University, Medford, 
Massachusetts 02155, USA}
\newcommand{\UCL}{Physics and Astronomy Department, University College 
London, 
Gower Street, London WC1E 6BT, United Kingdom}
\newcommand{\Virginia}{Department of Physics, University of Virginia, 
Charlottesville, Virginia 22904, USA}
\newcommand{\WSU}{Department of Mathematics, Statistics, and Physics,
 Wichita State University, 
Wichita, Kansas 67260, USA}
\newcommand{\WandM}{Department of Physics, William \& Mary, 
Williamsburg, Virginia 23187, USA}
\newcommand{\Wisconsin}{Department of Physics, University of 
Wisconsin-Madison, Madison, Wisconsin 53706, USA}
\newcommand{\deceased}{Deceased.}
\affiliation{\ANL}
\affiliation{\Atlantico}
\affiliation{\Bandirma}
\affiliation{\BHU}
\affiliation{\Caltech}
\affiliation{\CUSB}
\affiliation{\Charles}
\affiliation{\Cincinnati}
\affiliation{\CSU}
\affiliation{\CTU}
\affiliation{\Delhi}
\affiliation{\Erciyes}
\affiliation{\FNAL}
\affiliation{\FSU}
\affiliation{\UFG}
\affiliation{\Guwahati}
\affiliation{\Homi}
\affiliation{\Houston}
\affiliation{\Hyderabad}
\affiliation{\IHyderabad}
\affiliation{\IIT}
\affiliation{\Imperial}
\affiliation{\Indiana}
\affiliation{\ICS}
\affiliation{\INR}
\affiliation{\IOP}
\affiliation{\UIowa}
\affiliation{\ISU}
\affiliation{\Irvine}
\affiliation{\JINR}
\affiliation{\Magdalena}
\affiliation{\MSU}
\affiliation{\Duluth}
\affiliation{\Minnesota}
\affiliation{\Mississippi}
\affiliation{\NISER}
\affiliation{\OSU}
\affiliation{\Panjab}
\affiliation{\Pitt}
\affiliation{\QMU}
\affiliation{\SAlabama}
\affiliation{\Carolina}
\affiliation{\SMU}
\affiliation{\Sussex}
\affiliation{\Syracuse}
\affiliation{\Texas}
\affiliation{\Tufts}
\affiliation{\UCL}
\affiliation{\Virginia}
\affiliation{\WSU}
\affiliation{\WandM}
\affiliation{\Wisconsin}

\author{S.~Abubakar}
\affiliation{\Erciyes}

\author{M.~A.~Acero}
\affiliation{\Atlantico}

\author{B.~Acharya}
\affiliation{\Mississippi}

\author{P.~Adamson}
\affiliation{\FNAL}









\author{N.~Anfimov}
\affiliation{\JINR}


\author{A.~Antoshkin}
\affiliation{\JINR}


\author{E.~Arrieta-Diaz}
\affiliation{\Magdalena}

\author{L.~Asquith}
\affiliation{\Sussex}


\author{A.~Aurisano}
\affiliation{\Cincinnati}






\author{N.~Balashov}
\affiliation{\JINR}

% ...

\collaboration{The NOvA Collaboration}
\noaffiliation




\begin{abstract}
We report a search for a magnetic monopole component of the cosmic-ray flux in a 2743-live-day exposure of the NOvA experiment's Far Detector, a 14\,kt segmented liquid scintillator detector designed primarily to observe GeV-scale electron neutrinos.  No events consistent with monopoles were observed, setting an upper limit on the flux of $8\times 10^{-16}\,\mathrm{cm^{-2}s^{-1}sr^{-1}}$ at 90\% C.L. for monopole speed $6\times 10^{-4} < \beta < 5\times 10^{-3}$ and mass greater than $10^{9}$\,GeV.  Because of NOvA's small overburden of 3~meters-water equivalent, this constraint covers a previously unexplored low-mass region.
\end{abstract}

\maketitle


\section{Introduction}
\label{sec:introduction}

Magnetically charged particles were hypothesized by Dirac in 1931~\cite{Dirac:1931kp} and are generically predicted by grand unified theories (GUTs)~\cite{Tanabashi:2018oca,tHooft:1974kcl,Polyakov:1974ek}.  Although GUT-scale monopoles of $\sim$$10^{17}$--$10^{18}$\,GeV are often assumed, recent theoretical work suggests possible masses as light as $\sim$$10^7$\,GeV~\cite{Kephart:2001ix, Tanabashi:2018oca}.  Searches over the past century for a monopole component of the cosmic ray flux have yet to find convincing evidence~\cite{Groom:1986ps}. Stringent limits have been set on slow-moving ($\beta < 0.01$) GUT-scale monopoles by underground experiments~\cite{macro}.  Weaker limits exist for slow monopoles in the range $10^5\,\mathrm{GeV} < m < 10^{12}$\,GeV from mountaintop experiments~\cite{slim}. Monopoles have also been probed assuming their possible production in proton-proton collisions at the LHC~\cite{MoEDAL_first,MoEDAL_second,MoEDAL_third}. In this paper, we focus on the possibility that there is a flux of slow cosmic-ray magnetic monopoles.  As a large low-elevation surface detector, the NOvA Far Detector is sensitive to a combination of monopole masses and speeds not previously accessible.  This analysis was previously performed on a 95-live-day sample of NOvA data~\cite{nova_monopole} but this paper extends it to a 2743-live-day sample with updated analysis techniques.

Since the previous analysis, the detector gain has been increased, which makes the detector more sensitive to lower energy depositions.  This is particularly useful in detecting the slowest magnetic monopoles since their energy deposition rate falls off as $\beta$ decreases~\cite{nova_monopole}.


The \nova experiment is designed primarily to measure long-baseline neutrino oscillations~\cite{Acero:2019ksn}.  The measurement of these oscillations in neutrinos and antineutrinos gives information about the neutrino mass ordering, the PMNS matrix mixing parameters ($\theta_{23}$, $\Delta m^{2}_{32}$), and the CP-violating phase $\delta_{CP}$.  \nova uses two detectors, the 0.3\,kt Near Detector underground at Fermilab, 50\,km west of downtown Chicago, IL, and the 14\,kt Far Detector (FD) near Ash River, MN.  A beam of muon neutrinos produced at Fermilab travels through the Earth to the FD.

Due to its surface location, monopoles with \mbox{$m \gtrsim 10^9$\,GeV} can reach the FD without being absorbed by the atmosphere or overburden, while its size and trigger design allow identification of slow tracks.  Compared to a dedicated underground monopole detector, \nova is not optimized for monopole detection and must contend with a large cosmic muon background.  A detailed description of \nova's solid angle coverage and acceptance region is provided elsewhere~\cite{nova_fast_monopole}.

\begin{figure}
\includegraphics[width=\columnwidth]{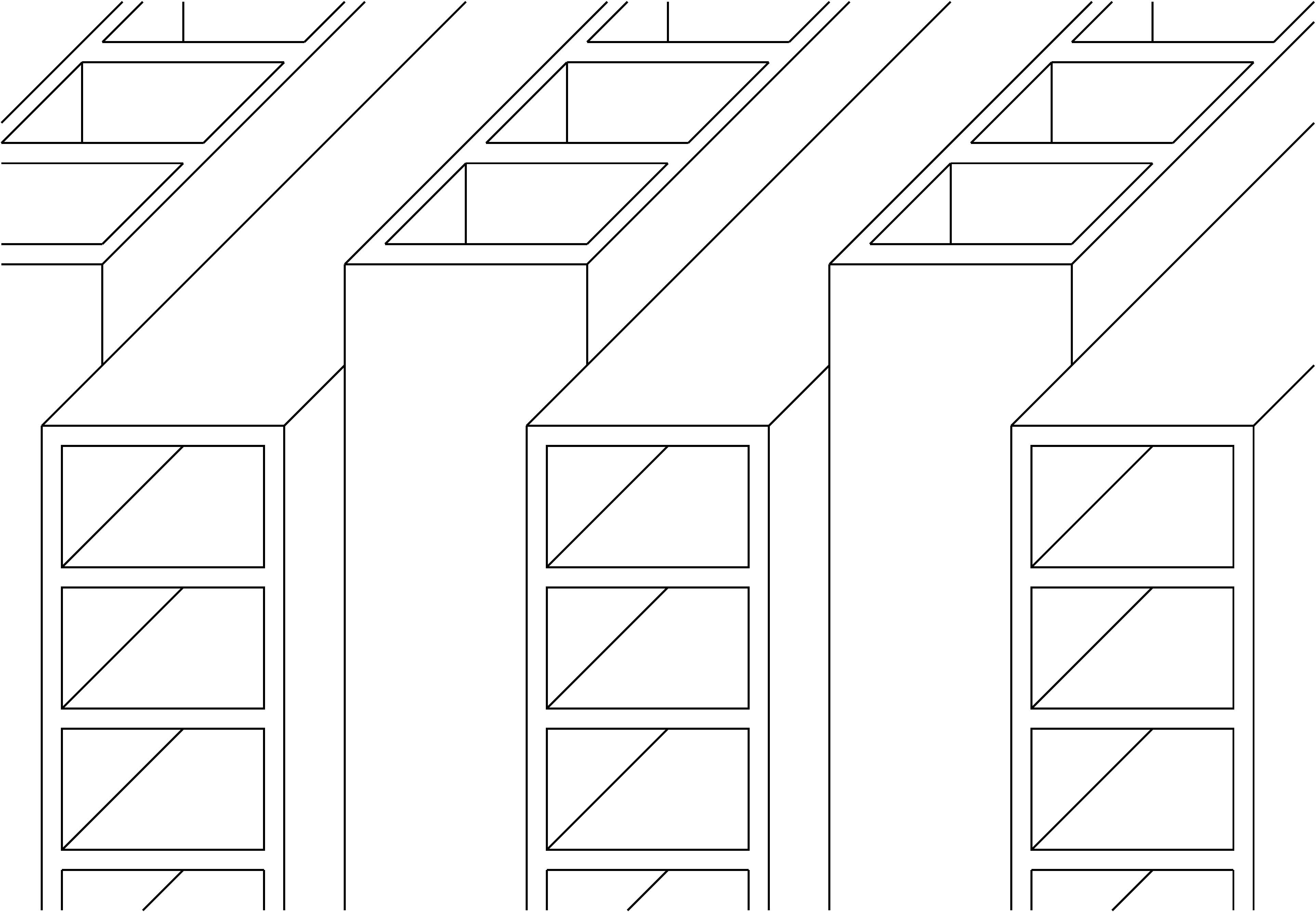}
\caption{Schematic of a corner of the NOvA detector.  The $z$ direction is to the right, perpendicular to the 15.5~meter-long cells, the ends of which are shown.}
\label{fig:det}
\end{figure}

The \nova FD is on the surface, with a concrete and barite overburden of 3~meters-water equivalent. It is a segmented detector with dimensions \unit[15.5]{m} by \unit[15.5]{m} by \unit[59.8]{m}, consisting of 896 planes of 384 plastic cells each filled with organic liquid scintillator. Each cell is \unit[15.5]{m} by \unit[4]{cm} by \unit[6]{cm}.  Planes alternate between $x$ and $y$ orientations (see Fig.~\ref{fig:det}), with signals acquired from two projected views, $xz$ and $yz$, separately.  The $x$-direction points $28^\circ$ south of west, $y$ is vertical, and $z$ is the long axis of the detector such that the three form a right-handed coordinate system.  Light produced in the cells by ionizing particles is collected by a loop of wavelength-shifting fiber and converted to electrical signals by avalanche photodiodes (APDs).  All APD signals are continuously digitized at 2\,MHz.  In each cell, samples that rise above a threshold defined to exclude the majority of noise are retained for further trigger processing.  Such a sample from one cell is called a ``hit.''  A cell can satisfy the criteria for a hit as often as once every three clock cycles, or 1.5\,$\mu$s.  Each hit provides a 2D position; 3D trajectories are reconstructed using hits from the two views~\cite{novatdr}.


The remainder of this paper is organized as follows. We state our assumptions about monopole interactions and our detector simulation, which are used to determine detection efficiency, in Sec.~\ref{sec:simulation}.  NOvA's dedicated monopole trigger is described in Sec.~\ref{sec:trigger}, the offline event selection is described in Sec.~\ref{sec:selection}, the background studies are reported in Sec.~\ref{sec:background}, and our measurement results are presented in Sec.~\ref{sec:results}.

\section{Simulation}\label{sec:simulation}

There is no prior knowledge on the distribution of the direction of the monopoles, so we 
use an isotropic distribution of monopole directions in our simulation.  We are interested only in the monopole's activity inside the detector, so we start by choosing a random surface point on the detector as the entry point for the monopole.  The two initial parameters that the monopole possesses before it enters the detector are its direction and its speed ($\beta$).  The speed determines the energy deposition rate~\cite{Groom:1986ps}.  The monopole energy is very large relative to the energy lost in the detector, so we assume that the monopole will traverse the entire detector and not stop.  Furthermore, the monopole is assumed to travel in a straight line.  The simulated monopole is then stepped through the detector using Geant4.  The amount of energy deposited in each cell depends on the path length of the monopole through that cell and is determined 
using the standard NOvA detector response simulation.  The simulation stops as soon as the monopole steps outside of the detector.  The details of the energy deposition model and its implementation in the simulation using a conservative assumption (90\% of the nominal Geant4 dE/dx value) are described in the previous search and were not modified for this one~\cite{nova_monopole}.

Each simulated monopole was combined with 5\,ms of zero bias data from the FD (i.e., data with typical running conditions, saved to permanent storage without regard to its content).  The trigger operates on data blocks of this duration. This results in an event that contains both the simulated monopole and real detector activity which is dominated by cosmic rays.  Samples at various monopole speeds [$10^{-4} \leq \beta \leq 10^{-2}$] were used to measure how well the search algorithm can identify slow monopoles and differentiate them from the cosmic-ray background.  Sixteen thousand simulated monopoles were generated for each $\beta$ point shown in Table~\ref{tab:limits}.

\section{Online Trigger Algorithm}
\label{sec:trigger}


Using NOvA's data-driven trigger system~\cite{Norman:2015ete}, the FD is able to isolate interesting physics signals among 150\,kHz of cosmic rays.  This trigger system operates entirely in software and is able to perform arbitrary analyses of incoming data, although the complexity is limited by CPU time. The event topology in this search is a straight track traversing the FD, in any direction, with a speed that is a small fraction of the speed of light.  The trigger was optimized for $\beta=10^{-3}$ magnetic monopoles -- the middle point of our initial [$10^{-4} \leq \beta \leq 10^{-2}$] search range.

\begin{figure}
\centerline{\includegraphics[width=\columnwidth]{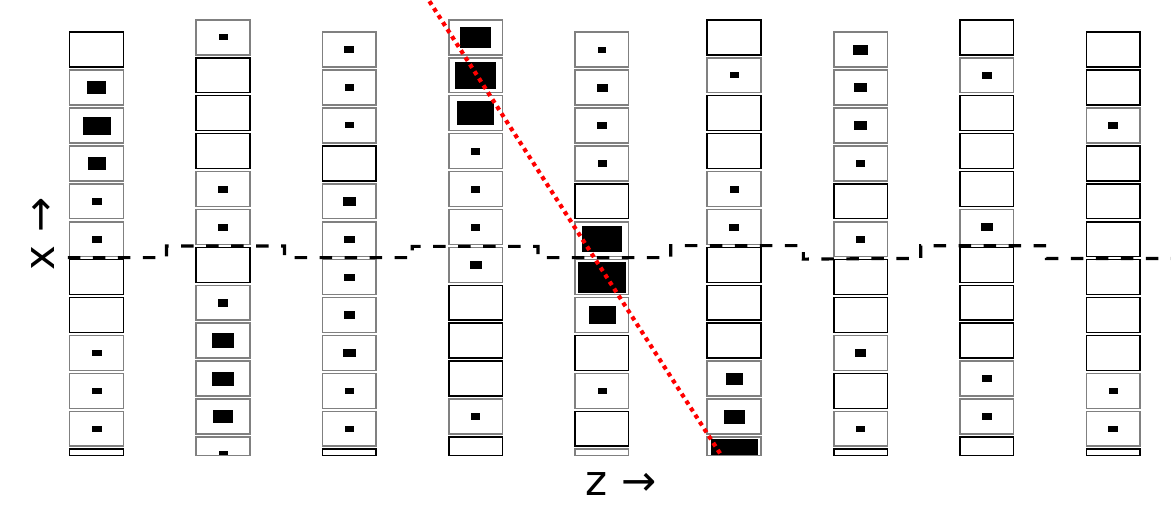}}
\caption{Hit selection in the trigger algorithm.  Cells near the $+x$ edge in the $xz$ view are shown.  Filled boxes represent hits in a 5\,ms window; size is proportional to signal strength.  The dotted red line shows the path of a simulated monopole; hits off this line are zero bias data.  Hits above the horizontal dashed line are considered to be on the detector edge.}
\label{fig:trigger_schematic}
\end{figure}

Pairs of hits within 2\,$\mu$s of each other and in neighboring $xz$ and $yz$ planes define 3D positions.  Each 3D pair that defines a position within six cells or five planes of the surface of the detector (see Fig.~\ref{fig:trigger_schematic}) is retained for further processing.  Using the $xz$ view alone for CPU efficiency, the trigger forms track seeds consisting of two selected hits on different detector faces and a time difference consistent with originating from a particle with $10^{-4.4} < \beta_{2D} < 10^{-2.3}$ in the $xz$ view.  The lower speed limit was set to approximately correspond to when the average monopole track would no longer be contained within a 5\,ms data block; CPU time was saved by not considering slower particles.  This 2D speed can correspond to a 3D speed as large as $\beta \approx 10^{-2}$.

For each track seed, the algorithm identifies hits that lie on a 20 cell (80\,cm) wide ``road'' with its center along the line connecting the seed hits.  It then looks for gaps between adjacent hits on the road and identifies the maximum plane gap (in the $z$-direction) and the maximum cell gap (in the $x$-direction).  If there is a plane gap larger than 30 planes (200\,cm) or a cell gap larger than 20 cells, the algorithm rejects the track seed.  Otherwise, a time window of data containing the track seed plus 4\,$\mu$s before and afterwards is written out to permanent storage.  Because a true monopole would generally produce many track seeds that pass the algorithm's selection, only every fourth track seed is checked, to save CPU time (during running periods I. and II.). Since a monopole that created many track seeds would also have many hits on its road, the check for gaps rejects background without significantly affecting the signal.

\begin{figure}
    \centering
    \includegraphics[width=\figurescale]{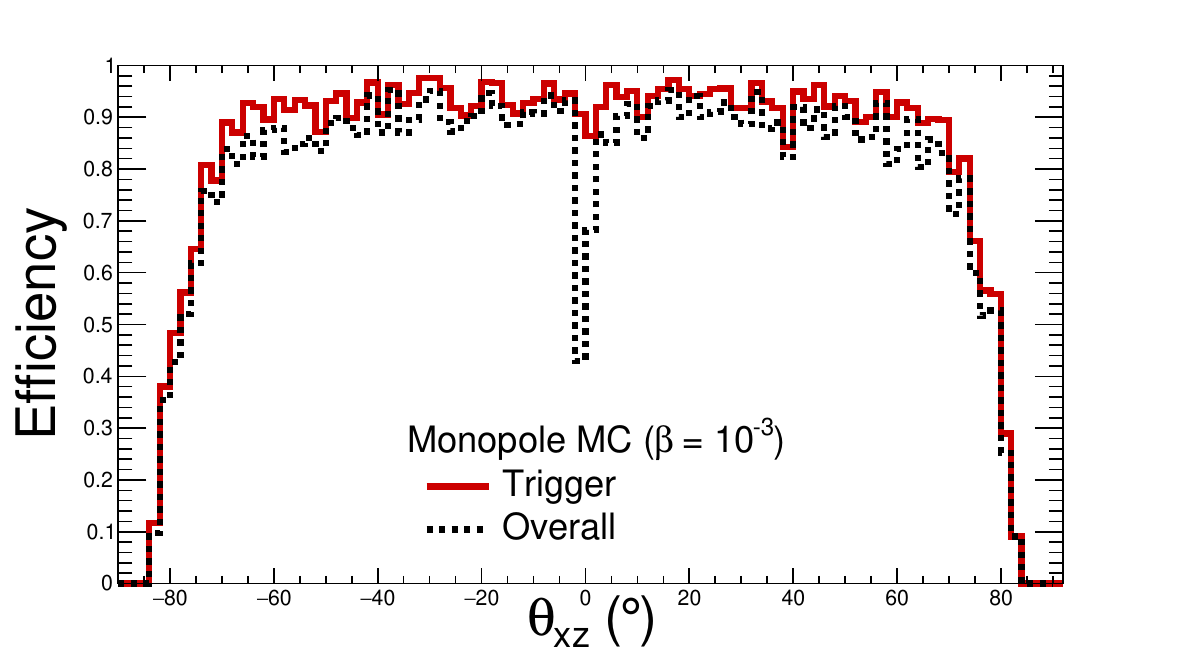}
    \caption{Trigger and overall selection efficiencies vs. $\theta_{xz}$ (2D angle measured from the $z$ axis in the $xz$ view) for monopoles that cross at least 10\,m of the detector with $\beta=10^{-3}$.  Efficiencies in the $yz$ view have the same form.  The drop in overall efficiency for nearly horizontal tracks ($\theta_{xz} \sim 0^\circ$) is due to the detector structure, which is explained in the text.}
    \label{fig:thetax}
\end{figure}

Integrated over all angles, the efficiency for triggering on a monopole with $\beta = 10^{-3}$ that intersects the detector with a true crossing length of at least 10\,m is 83\% (see Fig.~\ref{fig:thetax}).  $\theta_{xz}$ is the two-dimensional angle measured from the $z$ axis in the $xz$ view.  Most of the efficiency loss at the trigger level is caused by the need for the monopole to intersect enough cells in both the $xz$ and $yz$ views for its 3D pairs on the surface to be examined.  The overall efficiency drop for nearly horizontal tracks ($\theta_{xz} \sim 0^\circ$) in Fig.~\ref{fig:thetax} is due to the detector structure.  As can be seen in Fig.~\ref{fig:trigger_schematic}, the cell centers shift along the $x$ axis, so an almost horizontal track will have alternating $x$ coordinates affecting the track reconstruction by degrading the track's $r^2$ value, which is defined in Sec.~\ref{sec:selection}.  This effect is also present in the $yz$ projection.

This analysis considered data collected from February 5, 2016 through October 12, 2024 and while the trigger algorithm itself was not changed in this period, some configuration parameters were adjusted to adapt to the FD running conditions.  These parameter changes divide the data into the following three running periods:
\begin{enumerate}
\item[I.] February 5, 2016 to October 14, 2017: \\
A special time-out bit was added to mark events that exhausted the available processing time due to the high hit multiplicity of the event.  These events were then fully (i.e. all 5\,ms) written to permanent storage.  Detailed offline checks of these events revealed that most of them contained high energy showers but no track-like structures that we were searching for.

\item[II.] November 7, 2017 to September 16, 2021: \\
The timed-out events were no longer stored in their entirety, but instead only the time window that contained the current track seed plus a 4\,$\mu$s buffer before and after.

\item[III.] September 16, 2021 to October 12, 2024: \\
The upstream trigger algorithms that provide the hit collection to the monopole trigger were modified to remove noise and uncorrelated hits in an effort to improve our Supernova trigger sensitivity~\cite{nova_sn}.  This reduction in undesired hits allowed us to improve our monopole trigger by checking every track seed instead of only every fourth as in the previous two run periods.
\end{enumerate}

\section{Event Selection}
\label{sec:selection}


The initial stage of offline event selection is track reconstruction, first of speed-of-light tracks, then of slow tracks.  Speed-of-light tracks are primarily cosmic-ray muons.  Such a particle takes 50\,ns to traverse the height or width of the detector and 200\,ns to traverse the length; the hit timing resolution is typically 20\,ns.  Candidate tracks have hundreds of hits, making speed-of-light tracks easily distinguishable from the slow tracks of interest in this search. To find speed-of-light tracks, hits were clustered using their proximity in time and space. Within these clusters, straight lines consisting of several hits were identified and joined together to form tracks.  All hits belonging to such tracks were removed from consideration for monopole track reconstruction.  Since removed tracks can overlap a potential monopole track, some true monopole hits may be discarded at this step, reducing the search efficiency.  Our simulations showed fewer than 1\% of monopole hits would be discarded in this way across the range of $\beta$ considered.

Hits were then removed if they were separated from all other hits by at least two planes and two cells in their respective views.  This removal ensures that sparse tracks are not reconstructed out of stray hits arising from radioactive decays, low energy components of cosmic ray showers, hot channels, and other sources of background that are not reconstructed as speed-of-light tracks.  The remaining hits were reconstructed using the Hough tracking algorithm~\cite{hough} to identify straight line objects.  (Since the monopoles under investigation are so heavy, they do not undergo significant multiple scattering and should appear as perfectly straight lines up to the detector's resolution.) A line was fitted to each such collection of hits.  A candidate was required to meet the following three basic requirements, as well as others described below.  The track must (1)~have at least 20 hits in each view, (2)~cross at least 10 planes in each view, and (3)~have a reconstructed length of at least 10\,m.


The speed, linear correlation coefficient, and time gap fraction (defined below) were calculated for these candidate monopole tracks.  As the trigger algorithm searches for monopoles with $\beta < 0.01$, and the analysis strategy is optimized for slow monopoles, any candidate monopole track reconstructed with a speed above this was rejected, but is covered in~\cite{nova_fast_monopole}.

Since slow monopoles are not expected to be highly ionizing, distinguishing characteristics used in this search were the straightness of their tracks and their consistent slow speed.  In order to determine the speed, standard linear regression is applied to the space ($x$, $y$, or $z$) and time~($t$) values of the track hits.  This yields the speed in the $xt$, $yt$, and $zt$ projections separately and also the correlation coefficient ($r^2$), which is a measure of straightness.  A true monopole track would have $r^2$ close to unity.   The minimum of $r^2_{xt}$ and $r^2_{yt}$ is called $r^2_\mathrm{min}$.

A potential background was reconstruction failures in which two speed-of-light cosmic rays are identified as a single monopole track.  Such a background track would have a cluster of hits occurring early in time, a large time gap, and then another cluster of hits occurring later.  To remove such cases, the largest time gap between consecutive hits was required to be small, defined as follows.  The quantities $f_{xt}$ and $f_{yt}$ were calculated for each track.  For each view, $f$ is the ratio of the largest time gap between hits in the track to the total extent of the track in time.  A high-quality track has a value of $f$ close to zero, whereas a track built from two unrelated cosmic rays will have a value of $f$ close to unity.  The maximum of $f_{xt}$ and $f_{yt}$ is called $f_{\mathrm{max}}$.

The selection criteria  are unchanged from the previous search~\cite{nova_monopole} and are given below.  Before unblinding the full data sample, a small randomly selected sample (0.4\%), which was subsequently excluded from the analysis, was checked under the assumption that such samples would contain no monopoles and it was found that the following selection criteria still apply.  In addition to meeting requirements (1) through (3) from above, a candidate track must also meet the following requirements: (4) $\beta < 10^{-2}$, (5) $r^2_{\mathrm{min}} \ge 0.95$, and (6) $f_{\mathrm{max}} \le 0.2$.

\begin{figure}
\begin{center}
\includegraphics[width=\figurescale]{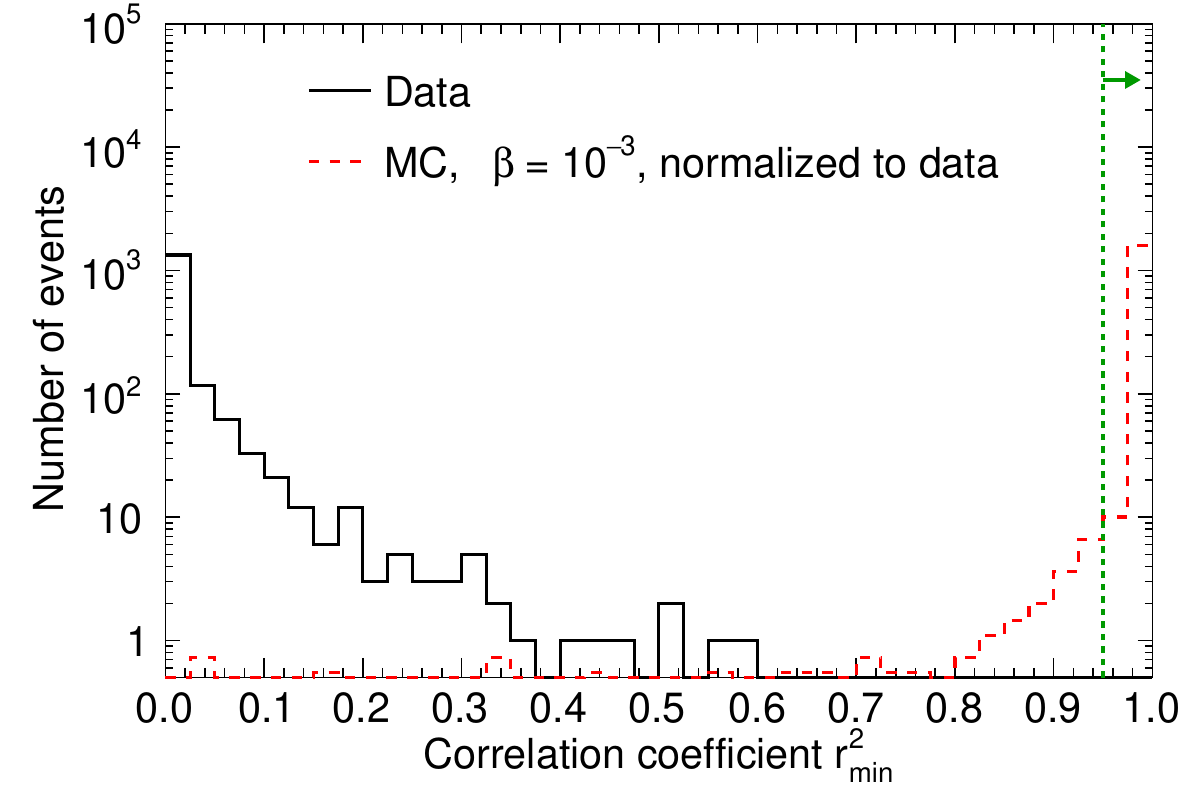}
\caption{Correlation coefficient $r^2_{\mathrm{min}}$ of monopole candidates.  Events shown pass selections 1--4 and 6.  Selection requirement 5 is displayed as a vertical dashed line. Monte Carlo simulation is normalized to the data for display purposes.}
\label{fig:r2_gap}
\end{center}
\end{figure}

Figure~\ref{fig:r2_gap} shows the strength of $r_\mathrm{min}^2$ in separating signal from background.  The cutoff values for these variables were not highly optimized, but rather chosen by hand to clearly lie far from the background while retaining most of the signal.


Finally, we planned to visually examine any event passing all selections to determine if it appeared to be an unanticipated background.

\begin{figure}
    \centering
    \includegraphics[width=\figurescale]{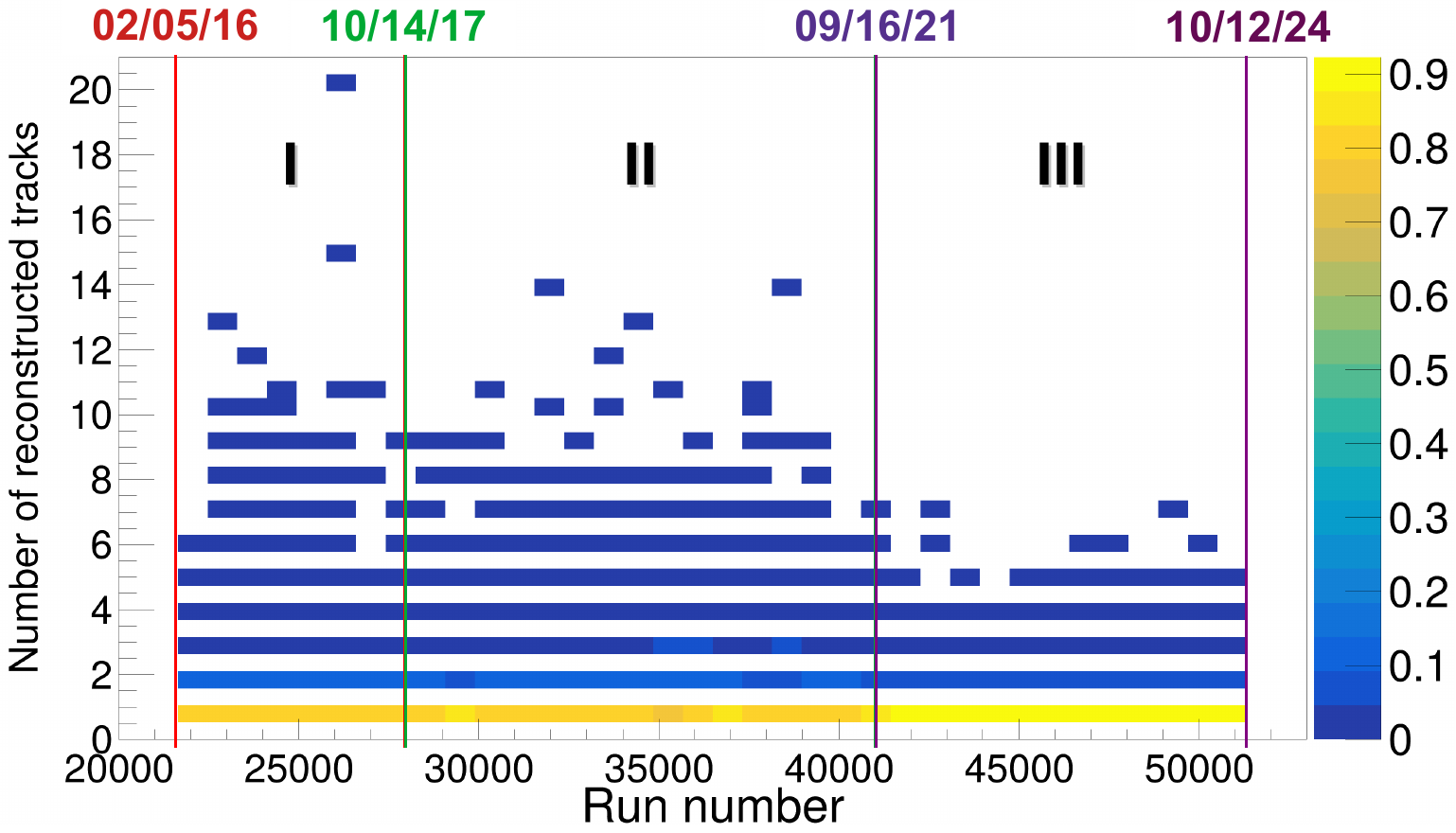}
    \caption{The number of reconstructed tracks vs. run number for the three run periods defined in Sec.~\ref{sec:trigger}.  The heat map indicates the fraction of events in each vertical bin, so it can be seen that the majority of reconstructed events contains only one track (yellow).  The reduced number of reconstructed tracks in the third running period is due to the improved noise and uncorrelated hit rejection.}
    \label{fig:Number_reco_tracks_vs_time}
\end{figure}

We store fully reconstructed events for only a subset of the data, which is the set of all triggered events having at least one reconstructed track as determined by the offline reconstruction software.  We monitored these files over time (number of reconstructed tracks and hits per run, length and angular distributions per reconstructed track, etc.) to ensure their integrity and quality and we found that the monitored quantities were all constant in time. Figure~\ref{fig:Number_reco_tracks_vs_time} is one of these quality plots that shows the number of reconstructed tracks vs. time (run number) over the full data sample.  This histogram shows the stability of our data with the majority of events having one reconstructed track.  This method reduced the overall data volume stored for this analysis by a factor of twenty, which allowed faster processing at the time of the final analysis.

\section{Background Studies}
\label{sec:background}

\begin{figure}
    \centering
    \includegraphics[width=\figurescale]{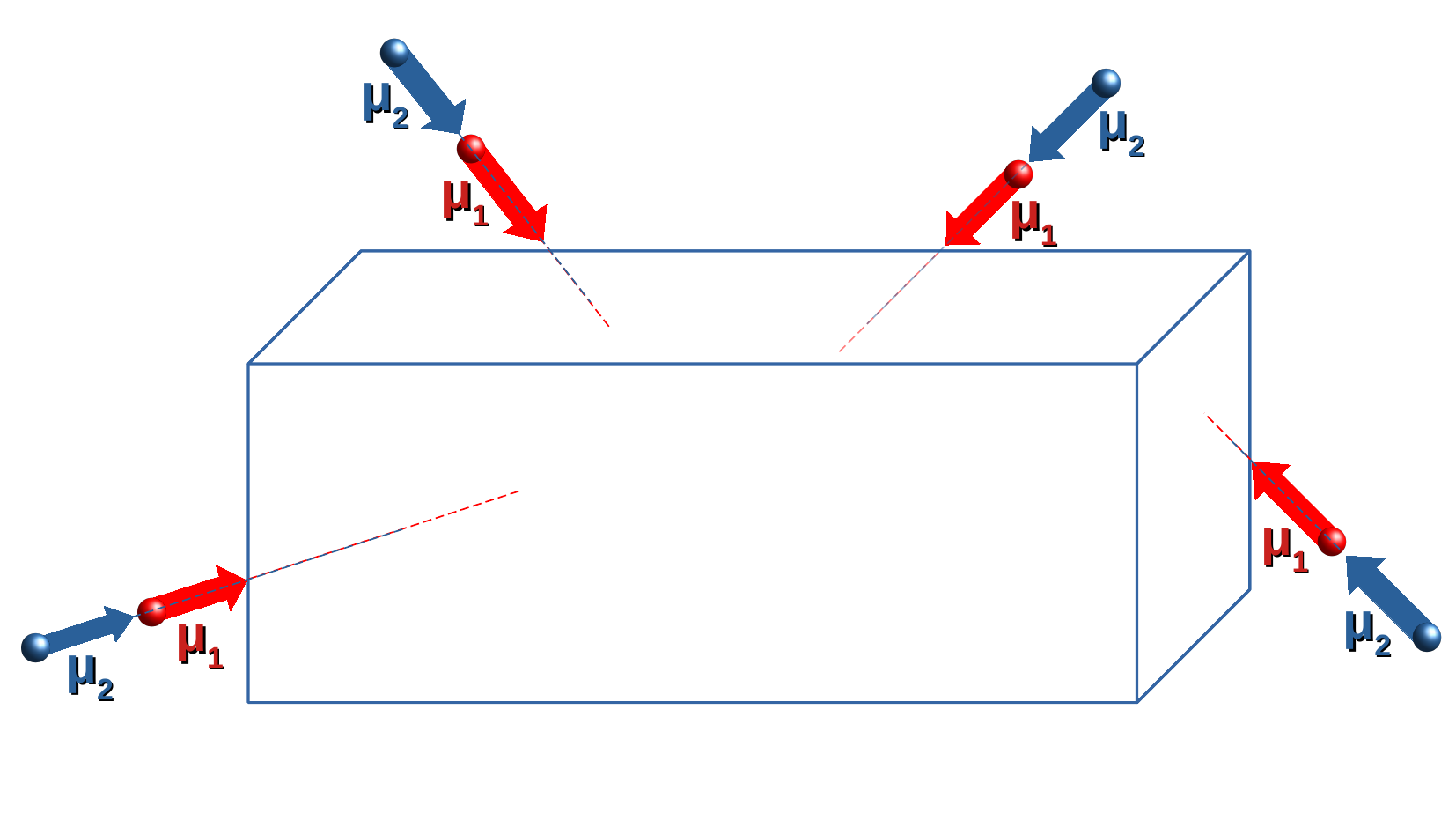}
    \caption{This schematic illustrates the general strategy of isotropically generated muon pairs with shifted arrival times.  Muon 1 would arrive first and muon 2 would then arrive along the same path at a later time.  The difference in arrival times was varied uniformly between 5 and 75~$\mu$s.  The indicated trajectories illustrate the isotropic nature of the simulation, but each simulated event contains only one muon pair.}
    \label{fig:muon_strategy}
\end{figure}

In the previous analysis~\cite{nova_monopole}, it was found that the most signal-like events were caused by two speed-of-light tracks in the same location at slightly different times.  We therefore performed a dedicated simulation of two such tracks incident on the FD from the same direction, but with a variety of time intervals separating their arrival times (see Fig.~\ref{fig:muon_strategy}).  We also varied the incident direction isotropically to check all directions with the exception of muons coming from below. Their momentum was chosen to yield muons traversing the entire detector without showering and was varied uniformly in the range from 15 to 25~GeV.  Their shifted time was chosen to coincide with the $\beta$ range of interest and was varied uniformly in the range from 5 to 75~$\mu$s.
In the end, we found that the reconstruction algorithm rejected all of these events and did not identify even a single reconstructed track candidate.  This improvement resulted from a new outlier removal method, which refined the position and time reconstruction of the track, making it less likely to be combined with another track.

The unlikeliness of such a background event occurring can also be considered statistically.  The probability for two muons to have the same trajectory in space but shifted in time to give a false-positive signal was evaluated assuming a Poisson distribution.  If we use the muon coincidence probability in space and time as the Poisson mean, we find that the probability of such an event occurring in the total running period considered here is less than $10^{-6}$.

\section{Results}\label{sec:results}

The data set for this search was recorded from February 5, 2016 through October 12, 2024. During this date range, the detector was live for 2985~days and provided good data for $2.55\times10^8$\,s.  The data-driven trigger system was 92.92\% efficient during this time period, where the inefficiency was caused by the available CPU time for the trigger to process incoming data being exhausted.  The corrected live time is therefore $2.37 \times 10^{8}$\,s (2742.8~days).

\begin{figure}
\begin{center}
\includegraphics[width=\figurescale]{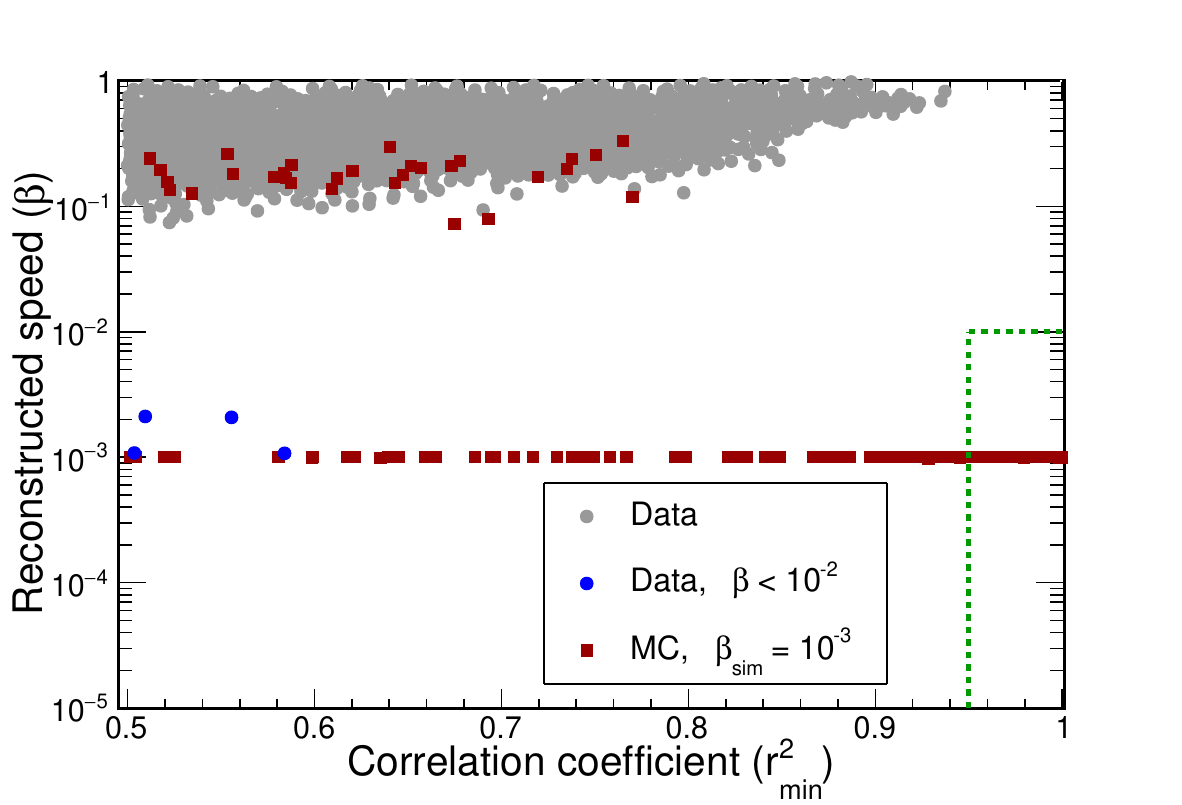}

\caption{Reconstructed monopole speed vs.\ $r^2_\mathrm{min}$ for events passing selections 1--3 and 6.  Data events are shown as grey circles.  The four data events that additionally meet selection criterion 4 ($\beta < 10^{-2}$) are highlighted in blue.  Simulation events for $\beta_{\mathrm{sim}} = 10^{-3}$ are shown as red squares.  Events must be in the dashed box in the lower right to be selected.}
\label{fig:scatter}
\end{center}
\end{figure}

All of the triggered data events were recorded and reconstructed if the event contained at least one reconstructed track as described in Sec.~\ref{sec:selection}.  The final data set contained 9,300,025 events, none of which fell into the signal region.  Figure~\ref{fig:scatter} shows the distribution of the full sample in speed vs. $r_\mathrm{min}^2$. All data events are far from the signal region.  The large cluster of background events in the data around $\beta = 0.5$ was caused by speed-of-light muons not removed in the first reconstruction step.  The four data events at low $r_\mathrm{min}^2$ near $\beta = 10^{-3}$ were caused by high-energy showers that deposited large amounts of energy.  In these rare cases the quiescent pedestal can droop and the front-end electronics will record a hit as the signals return to baseline at a predictable time after the actual energy deposit.  This tends to form rectangular clusters of spurious hits rather than lines, which explains the low $r^2_{\mathrm{min}}$ values.  

In the absence of any candidates in the signal region, the 90\% C.L. flux upper limit is $\Phi_{90\%} = 2.3/L$, where $L = \Omega \epsilon A t$ is the integrated product of acceptance and livetime, $\Omega$ is the solid angle coverage, $\epsilon$ is the efficiency, $A$ is the projected surface area of the FD visible to the monopole, and $t$ is the integrated livetime.  Each quantity is detailed below.

\begin{figure}
\centerline{\includegraphics[width=\figurescale]{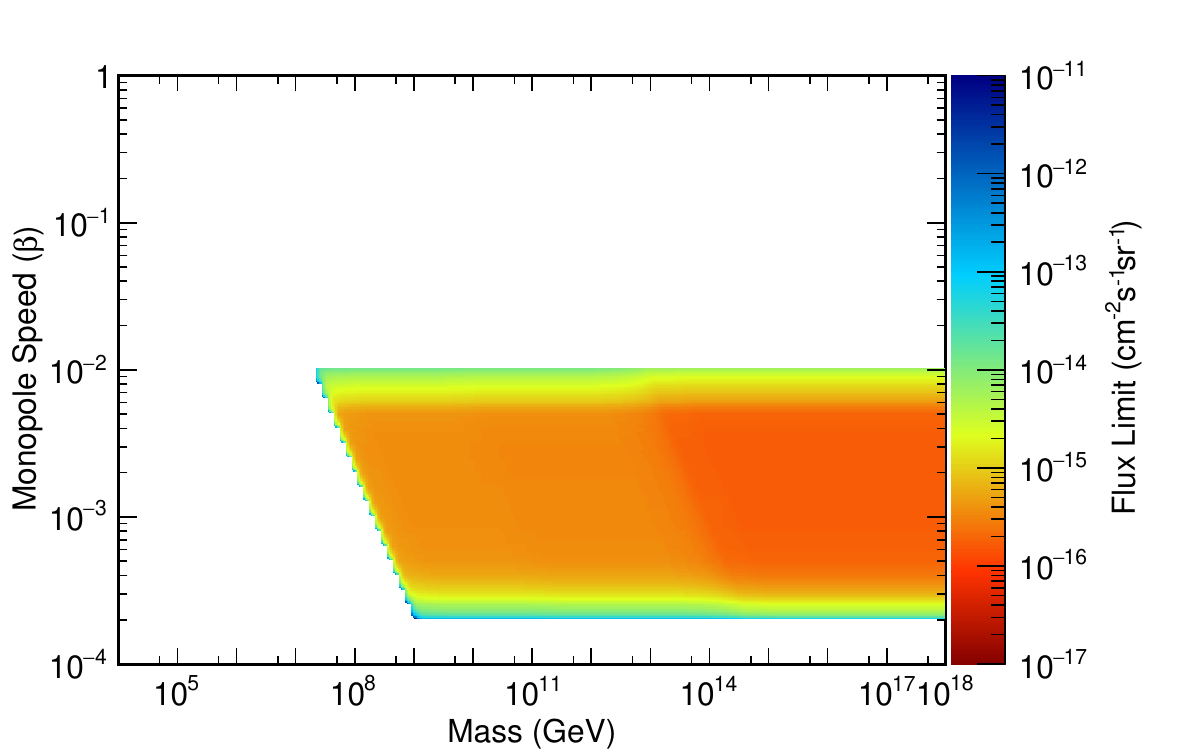}}
\caption{90\% C.L. upper limits on the magnetic monopole flux vs. monopole speed and mass.  Monopoles with masses above $10^{16}$~GeV possess enough kinetic energy to traverse the entire Earth yielding a $4\pi$ coverage while monopoles with lower masses down to $10^9$~GeV can traverse only the atmosphere and overburden.  This applies for all shown speeds, but the plot shows that NOvA is also sensitive to lower mass monopoles with higher speeds since they would have more kinetic energy.}
\label{fig:coverage}
\end{figure}

Limits are reported for the two major coverage scenarios: partial coverage where $\Omega = 1 \pi$, and full coverage where $\Omega = 4 \pi$. The coverage depends on the kinetic energy of the monopole, which is calculated from the monopole's speed and mass.  If the monopole's energy is sufficient, it could traverse the entire planet.  In this case, the FD has $4\pi$ coverage.  The $1 \pi$ coverage 
is the approximate extent of the solid angle around the zenith that does not go through either the whole Earth or the thicker part of the overburden around the detector.
Figure~\ref{fig:coverage} shows the upper limits of the magnetic monopole flux depending on monopole mass and speed.

We calculate the detector's projected area, $A$, for each simulated monopole's trajectory.  The reconstruction efficiency also depends on this trajectory, so for each monopole speed, the product was determined event by event, $\epsilon A \equiv \langle \epsilon_i A_i \rangle$.  The overall efficiency, considering both trigger efficiency and analysis selection, was 81\% (see Fig.~\ref{fig:thetax}).  Detector-based systematic uncertainties were negligible.  We considered uncertainties in livetime and solid angle and found that each was well under 1\%.


Table~\ref{tab:limits} shows flux limits vs. $\beta$.  Limits are shown for two mass values, $10^9$\,GeV and $10^{16}$\,GeV.  The former is the smallest mass for a monopole that would reach the detector through the atmosphere and detector overburden alone at the lower limit of the range of speeds considered, $\beta = 10^{-3.6}$. The latter is the smallest mass that would reach the detector through the Earth at $\beta = 10^{-3.6}$.  These values are shown in Fig.~\ref{fig:coverage} vs. monopole speed and mass.

\begin{figure}
\centerline{\includegraphics[width=\figurescale]{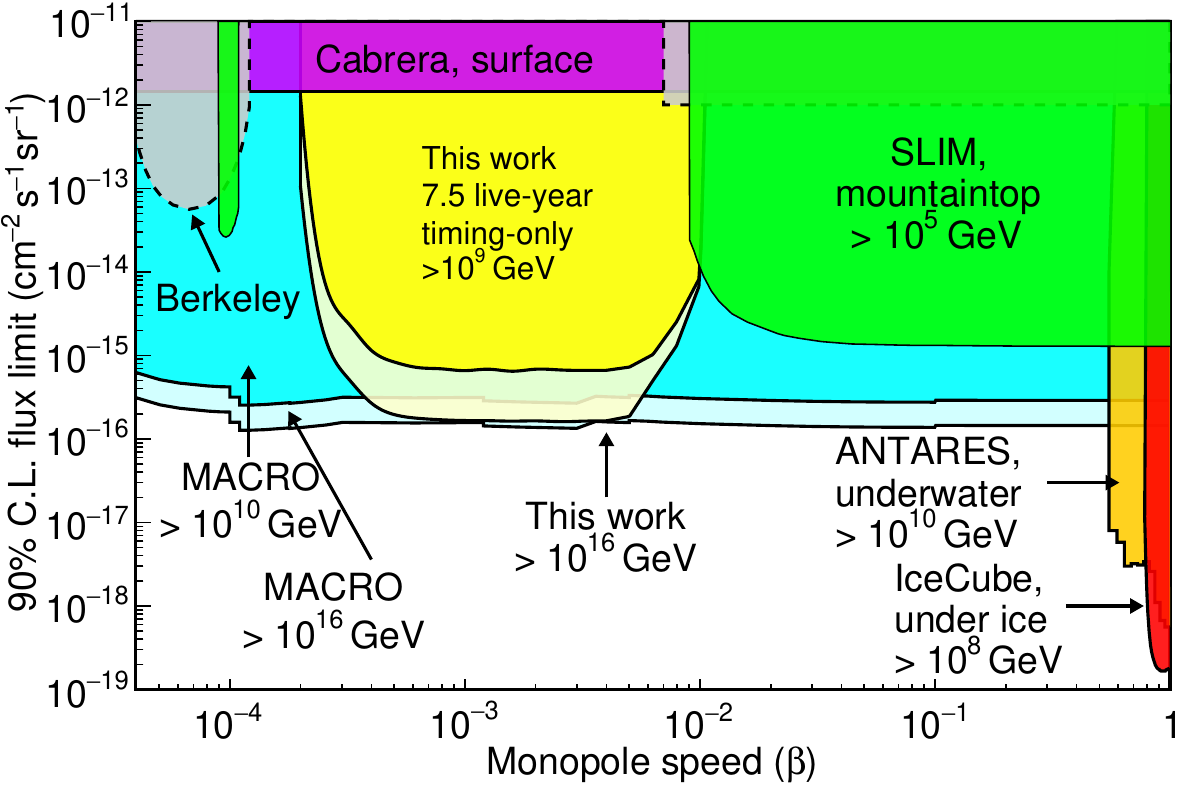}}
\caption{Upper limits on the magnetic monopole flux. Results are shown for experiments~\cite{macro,slim,icecubefast} that do not assume proton decay catalysis~\cite{icecubeslow} nor ultrarelativistic monopoles~\cite{rice}.  In each region of speed-flux space, the experiment with the best mass reach is shown.  NOvA sets the only limits around \mbox{$\beta = 10^{-3}$} for monopoles lighter than $10^{10}$\,GeV.}
\label{fig:limit_plot}
\end{figure}

Sensitivity falls off at low $\beta$ as monopole energy deposition drops below the analysis threshold, given the assumption of a monopole with a single Dirac unit $g$ of charge.  This assumption is common to the previous experiments whose limits are displayed in Fig.~\ref{fig:limit_plot}.  At high $\beta$, the sensitivity was limited by the trigger design, but is covered in~\cite{nova_fast_monopole} using a different ionization-based trigger and offline analysis.

\begin{table}
{
\setlength{\tabcolsep}{9pt}
\caption{90\% C.L. upper limits on the magnetic monopole flux, in units of $10^{-15}\,\mathrm{cm^{-2}s^{-1}sr^{-1}}$.}\begin{center}
\begin{tabular}{c c c}
\hline
\hline 
$\beta$ & $m > 10^9$\,GeV & $m > 10^{16}$\,GeV \\
\hline
  $10^{-3.6}$ & \phantom{0}8.23 & \phantom{0}1.80\phantom{.0} \\
  $10^{-3.5}$ & \phantom{0}2.48 & \phantom{0}0.54\phantom{.0} \\
  $10^{-3.4}$ & \phantom{0}1.16 & \phantom{0}0.27\phantom{.0} \\
  $10^{-3.3}$ & \phantom{0}0.83 & \phantom{0}0.20\phantom{.0} \\
  $10^{-3.2}$ & \phantom{0}0.72 & \phantom{0}0.17\phantom{.0} \\
  $10^{-3.1}$ & \phantom{0}0.69 & \phantom{0}0.17\phantom{.0} \\
  $10^{-3.0}$ & \phantom{0}0.66 & \phantom{0}0.17\phantom{.0} \\
  $10^{-2.9}$ & \phantom{0}0.68 & \phantom{0}0.17\phantom{.0} \\
  $10^{-2.8}$ & \phantom{0}0.65 & \phantom{0}0.16\phantom{.0} \\
  $10^{-2.7}$ & \phantom{0}0.68 & \phantom{0}0.16\phantom{.0} \\
  $10^{-2.6}$ & \phantom{0}0.67 & \phantom{0}0.16\phantom{.0} \\
  $10^{-2.5}$ & \phantom{0}0.66 & \phantom{0}0.16\phantom{.0} \\
  $10^{-2.4}$ & \phantom{0}0.66 & \phantom{0}0.16\phantom{.0} \\
  $10^{-2.3}$ & \phantom{0}0.73 & \phantom{0}0.19\phantom{.0} \\
  $10^{-2.2}$ & \phantom{0}1.03 & \phantom{0}0.50\phantom{.0} \\
  $10^{-2.1}$ & \phantom{0}2.52 & \phantom{0}1.30\phantom{.0} \\
  $10^{-2.0}$ & \phantom{0}8.69 & \phantom{0}7.21\phantom{.0} \\
\hline\hline
\end{tabular}
\end{center}
\label{tab:limits}
}
\end{table}

\section{Conclusion}

By virtue of being a large segmented detector on the Earth's surface, the NOvA FD is uniquely sensitive to slow monopoles in the mass range below $10^{10}$\,GeV, which would not have reached previous detectors such as MACRO~\cite{macro}. We have constrained the flux of this population of monopoles in a large region of speed-mass space which has previously been unconstrained, setting an upper limit on the flux of $8\times 10^{-16}\,\mathrm{cm^{-2}s^{-1}sr^{-1}}$ at 90\%~C.L. for $6\times 10^{-4} < \beta < 5 \times 10^{-3}$ and mass greater than $10^{9}$\,GeV.  The NOvA Far Detector continues to be in operation collecting data using the trigger described~herein.  

\section*{Acknowledgements}

This document was prepared by the NOvA collaboration using the resources of the Fermi National Accelerator Laboratory (Fermilab), a U.S. Department of Energy, Office of Science, HEP User Facility. Fermilab is managed by Fermi Forward Discovery Group, LLC, acting under Contract No. 89243024CSC000002.  This work was supported by the U.S. Department of Energy; the U.S. National Science Foundation; the Department of Science and Technology, India; the European Research Council; the MSMT CR, GA UK, Czech Republic; the RAS, the Ministry of Science and Higher Education, and RFBR, Russia; CNPq and FAPEG, Brazil; UKRI, STFC and the Royal Society, United Kingdom; and the state and University of Minnesota.  We are grateful for the contributions of the staffs of the University of Minnesota at the Ash River Laboratory, and of Fermilab. For the purpose of open access, the author has applied a Creative Commons Attribution (CC BY) license to any Author Accepted Manuscript version arising.

\bibliographystyle{apsrev4-1}
\bibliography{monopole}
 
\end{document}